\def\nino#1{\tilde{\chi}_{#1}^{0}}
\def\cinop#1{\tilde{\chi}_{#1}^{+}}
\def\cinom#1{\tilde{\chi}_{#1}^{-}}

\documentclass[slac_one]{revtex4}
\usepackage{graphicx}
\usepackage{fancyhdr}
\begin{document}

\flushright{CLNS 05/1924}
\flushleft

\title{{\small{2005 International Linear Collider Workshop - Stanford,
U.S.A.}}\\ 
\vspace{12pt}
Measuring Mass and Cross Section Parameters at a Focus Point Region} 

%

\author{R. Gray\footnote{This talk was given by R. Gray, describing ongoing 
work performed in collaboration with the other authors.}, J. Alexander, K. M. Ecklund, L. Fields, D. Hertz,
C. D. Jones, J. Pivarski}
\affiliation{Cornell University, Ithaca, NY, 14850, USA}
\author{A. Birkedal, K. Matchev}
\affiliation{University of Florida, Gainesville, FL, 32603, USA}

\begin{abstract}

The purpose of this study is to determine the experimental uncertainties in
measuring mass and cross section parameters of SUSY particles at a 500 GeV
Linear Collider. In this study SUSY is a point 
in the focus point region of mSUGRA parameter space that is compatible
with WMAP constraints on dark matter relic density.
At this study point the masses of the squarks and sleptons are very heavy, 
and the 
only SUSY particles accessible at the Linear Collider would be the three
lightest neutralinos, and the two lightest charginos: $\nino{1}$, $\nino{2}$,
$\nino{3}$, $\cinop{1}$, $\cinop{2}$, where $\nino{1}$ is the lightest 
supersymmetric particle (LSP).
The charginos or neutralinos may be pair produced, and the subsequent
decay cascades to the LSP allow us to measure the SUSY couplings and mass
spectrum. 
We find that by looking for the signature 2 jets plus 2 leptons plus
 missing energy we
can determine the mass of the LSP to within $1$ GeV uncertainty and that the
mass differences of $\nino{2}$ and $\nino{3}$ with the LSP mass can be 
determined to better than $0.5$ GeV. 

\end{abstract}

\maketitle

\thispagestyle{fancy}


\section{INTRODUCTION} 
This study represents work in an effort to establish a connection
between measurements at a future linear collider and cosmological 
data.
We assume that the LHC has run and discovered SUSY and that a 500 GeV
linear collider has run to make precision measurements. 
The ultimate goal of this study is
to determine whether or not the Linear Collider would be able to make precise
enough measurements of mass and cross sections of SUSY particles in order to
determine if the relic density of the LSP
would be consistent with WMAP constraints \cite{Spergel:2003cb}, 
and that the SUSY LSP actually
is the dark matter.

We have chosen the scenario where the discovered SUSY is mSUGRA in
 a focus point region \cite{matchev} with $M_{1/2}=300$ GeV, $A_0=0$,
$\tan\beta=10$, $\mu > 0 $, $m_t = 175$ GeV, $m_o = 3280$ GeV. In this focus 
point region the masses of the squarks and sleptons are very heavy, and the 
only SUSY particles accessible at the Linear Collider would be the three
lightest neutralinos, and the two lightest charginos: $\nino{1}$, $\nino{2}$,
$\nino{3}$, $\cinop{1}$, $\cinop{2}$, where neutralino $\nino{1}$ is the LSP.
The signed mass parameters and the production cross sections of the charginos
and neutralinos at this focus point region are given in tables \ref{massTable}, and \ref{crossTable}. Note that the mass parameters are signed. The sign of the
mass parameter will affect the decay distributions but not the kinematics.

This paper represents preliminary results on 
the expected experimental uncertainties 
in determining the mass parameters and production cross sections of SUSY 
particles at the linear collider. Specifically, we will discuss
 the 2 jet plus 2 lepton plus missing energy signature of $\nino{2}\nino{3}$
production.

\begin{table}[t]
\begin{center}
\caption{Neutralino and Chargino Mass Parameters}
\begin{tabular}{|l|c|c|}
\hline \textbf{Particle} & \textbf{Mass (GeV)} 
\\
\hline $\nino{1}$ & -107.7 \\
\hline $\nino{2}$ & -166.3 \\
\hline $\nino{3}$ & +190.0 \\
\hline $\cinop{1}$& -159.4 \\
\hline $\cinop{2}$& -286.6 \\
\hline
\end{tabular}
\label{massTable}
\end{center}
\end{table}

\begin{table}[t]
\begin{center}
\caption{Production Cross Sections of SUSY particle pairs in collisions
of 95\% left or right polarized electrons on unpolarized positrons.}
\begin{tabular}{|l|c|c|c|}
\hline \textbf{pair} &
\textbf{$\sigma_L$}(fb)& \textbf{$\sigma_R$}(fb)
\\
\hline $\cinop{1}\cinom{1}$& 940 & 119 \\
\hline $\cinop{1}\cinom{2}$& 48.9 & 40.3 \\
\hline $\nino{1}\nino{3}$ & 56.8 & 44.1 \\
\hline $\nino{2}\nino{3}$ & 92.4 & 70.9 \\
\hline
\end{tabular}
\label{crossTable}
\end{center}
\end{table}

\section{SIMULATION AND ANALYSIS CUTS}

\subsection{SUSY Signature}
We will analyze the production of $\nino{2}\nino{3}$. 
Since the squarks and sleptons are so heavy at this point in the 
parameter space, the dominant means for the $\nino{2}$ and $\nino{3}$
to decay is a transition to $\nino{1}$ (the LSP) by radiating a $Z$. 
Since the mass differences between $\nino{2}$, $\nino{3}$ and $\nino{1}$ are
all smaller than the $Z$ mass, the $Z$ is virtual.
In a detector, the two LSPs produced 
in the decay of the $\nino{2}\nino{3}$ pair will be seen only as 
missing energy. The two $Z$s produced in the decay will each
create one
of the following: $l\bar{l}$, $\nu\bar{\nu}$, $q\bar{q}$. A $l\bar{l}$ 
pair will
be visible as tracks, a $\nu\bar{\nu}$ will contribute to the missing
energy, and a $q\bar{q}$ will create at least 2 jets. Therefore, we will
be looking for one of the following detector signatures for $\nino{2}\nino{3}$
production: $2\textrm{Jet}+E_{\textrm{miss}}$, 
$2\textrm{Lepton}+E_{\textrm{miss}}$, 
$4\textrm{Jet}+E_{\textrm{miss}}$, $4{\textrm{Lepton}}+E_{\textrm{miss}}$, or 
$2{\textrm{Lepton}}+2{\textrm{Jet}}+E_{\textrm{miss}}$. 
For $\nino{2}\nino{3}$ production
we will use the mode $2\textrm{Lepton}+2\textrm{Jet}+E_{\textrm{miss}}$ 
because it has 
the smallest backgrounds from 
the other SUSY and Standard Model processes.

\subsection{Event Generation}

SUSY events were generated at Cornell using Isajet 7.69 \cite{isajetref}. 
This included
production via $e^+e^-$ collisions of the 
following neutralino and chargino pairs: $\cinop{1}\cinom{1}$, 
$\cinop{1}\cinom{2}$, $\nino{1}\nino{3}$, $\nino{2}\nino{3}$. The Standard
Model backgrounds
for this analysis were generated at SLAC using WHiZaRD 1.22 \cite{whizardref}. The background
include contributions from the following collisions: $e^+e^-$, $e^-\gamma$,
$e^+\gamma$, $\gamma\gamma$. The dominant backgrounds for this analysis 
generated by these collisions are $t\bar{t}$, $W^+W^-$ and $Z\bar{Z}$.
The beam in these event generators is a $\pm 95\%$ polarized 
electron beam on an unpolarized positron beam. The generators include
effects from initial state radiation (Beamstrahlung and Bremsstrahlung) which
can reduce the energy available for SUSY production.

\subsection{Detector Simulation and Reconstruction}

The detector simulation was done using LCD Root FastMC, SD Mar01, which is a 
root based fast Monte Carlo program produced at SLAC. 
Once we have tracks and showers 
the first task is to identify leptons that are isolated from jets and are 
candidates for having come directly from a neutralino or chargino decay. 
It is assumed that at the linear collider identification of electrons and 
muons will be near perfect and so lepton ID uses the Monte Carlo truth 
information. These leptons are  candidates
 for direct production from a SUSY decay if the lepton
momentum is greater than 10 GeV and the total energy from other particles
within 20 degrees of the lepton is less than 2 GeV. After these isolated
candidate leptons are identified, perfect track-shower matching is employed
to approximate the ``energy flow'' algorithms that would be used at
a real detector
and the remaining 4-vectors are 
divided into jets using the Durham algorithm with a $y_{cut}$ value
of $0.004$. 

We will use b-tagging in this analysis to help identify $t\bar{t}$ backgrounds.
Since the detector simulation used in this study is not advanced enough for
an honest b-tag, Monte Carlo truth information was used. 
The jets found by the Durham algorithm were matched to 
the generator level jets by an energy weighted comparison of the lists
of stable particles in the 
``found jet'' and the list of particles in the
generator level jet. A match quality
is defined by equation \ref{matchqual} where $E_i$ is the energy of particles
in list $i$, $E_{iaj}$ is the energy of particles that are listed in 
both list $i$ and list $j$, and $E_{inj}$ is the energy of particles listed
in list $i$ but not in list $j$. If the match quality is $1$, the two lists
are the same, if the match quality is $-1$ they have nothing in common. 
The generator level jet with the highest match quality for a 
``found jet'' is tagged to that found jet. 
If the ``found jet'' is matched to a jet that came from 
a $b$ quark, then we say it has a 50\% chance of being b-tagged.  

\begin{equation}
Q_{ij}=\frac{2 E_{iaj} - E_{inj} -E_{jni}}{E_i + E_j}
\label{matchqual}
\end{equation}

\subsection{Analysis Cuts for $\nino{2}\nino{3}\rightarrow l\bar{l}q\bar{q} \nino{1}\nino{1}$}

The analysis cuts for the 2 jet plus 2 lepton plus
 missing energy signal are as follows. There must be two leptons of 
opposite charge and same flavor that both passed
our isolation cuts. There must be 2 or 3 jets found in addition to the isolated
leptons. The missing energy must be greater than 275GeV. The transverse
momentum must be greater than 10GeV. The cosine of each track and jet
with the beam axis must be less than 0.95. The energy of each jet and letpon
must be less than 110GeV. Each jet is then anti-btagged, or the event is 
cut if either jet appears to have come from a b quark. This anti-btag cut
removes backgrounds from $t\bar{t}$. 

\section{MASS PARAMETER AND CROSS SECTION MEASUREMENT}

\subsection{Cross Section Measurement}

The cross section will simply be measured by counting the number of events
that pass our cuts with backgrounds subtracted, divided by the efficiency 
for passing those cuts, and divided by the luminosity.
With 500 $\textrm{fb}^{-1}$ of luminosity divided evenly
between left and right polarized beams we find statistical uncertainties on
polarized cross sections of
4\% and 5\% respectively. 

\subsection{Mass Parameters from Invariant Mass Distribution}

The mass difference between $\nino{2}$ and $\nino{1}$, and the mass
difference between $\nino{3}$ and $\nino{1}$, may be determined
by fitting the invariant mass distribution of the two leptons in the
2 letpon plus 2 or 3 jet data. For any decay $X_i \rightarrow l\bar{l}X_f$
the invariant mass of the two leptons will have a maximum possible value
of the mass difference between $X_i$ and $X_f$. Since $\nino{2}$ and $\nino{3}$
have different masses we will see two edges in the mass distribution.
One edge will be for the $\nino{2}$ and the other for the $\nino{3}$. 
In order to use the whole distribution, rather than just the edges, in our
mass determination, we will fit all the data to the expected invariant
mass distribution. This expected mass distribution will have very weak
dependence on the LSP ($\nino{1}$) mass, but will depend on the relative
sign of the $\nino{2}$ and $\nino{3}$ mass parameters with the $\nino{1}$
mass parameter. You can see
from table \ref{massTable} that $\nino{2}$ and $\nino{3}$ mass parameters
have different relative signs. We find that the efficiency for measuring the di-lepton is mostly flat 
in invariant mass beyond
10 GeV, so we will make no efficiency corrections to the 
distribution prior to performing the fit.

\begin{figure*}[t]
\centering
\includegraphics[width=135mm]{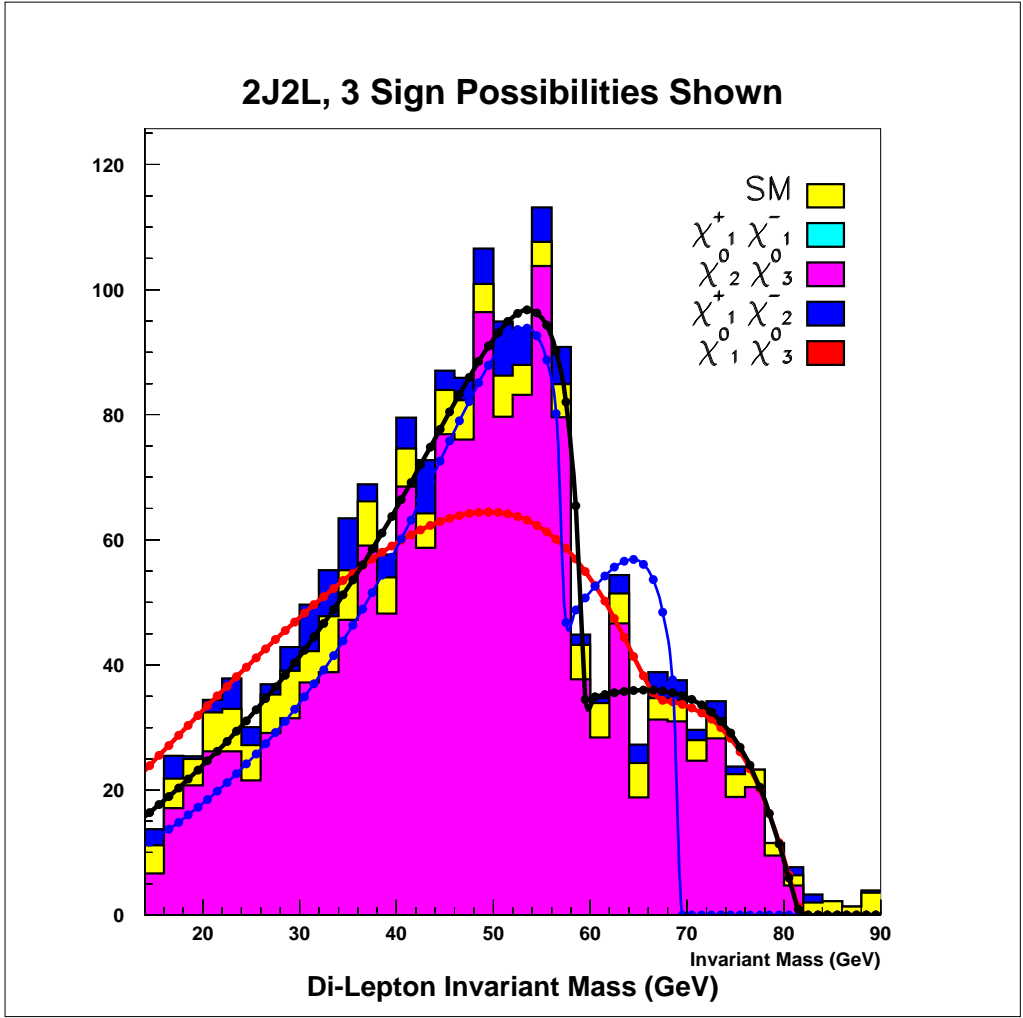}
\caption{This shows the invariant mass distribution of 
combined 250$\textrm{fb}^{-1}$ left polarized and 250 $\textrm{fb}^{-1}$ right
polarized data. The magenta is the signal. The yellow is
standard model backgrounds, and the blue is background
from chargino production. The three fits are for the 
different sign assumptions for the neutralino mass 
parameters. The red line is for (-,-), the blue line is for
(+,+), and the black line is for (+,-). The correct sign assumption is (+,-).
} \label{invMassDist}
\end{figure*}

The resulting invariant mass distribution can be seen in figure 
\ref{invMassDist}. We expect two edges in the distribution. 
We expect one at $58.6$ GeV for $\nino{2}$, and one at $82.3$ GeV
for the $\nino{3}$. Both edges are visible. 

The mass distribution of the background is fit to
an order 2 polynomial. The fit to the background
 distribution is combined with
the theoretical prediction for the SUSY component and an
unbinned log likelihood fit is performed. The fit is done
three times for the three possible combinations of relative
signs of $\nino{2}\nino{3}$ to the $\nino{1}$: (-,-), (+,+),
(+,-). We find that the possible sign combinations are easily
distinguished in the data, and that the best fit for
the correct sign combination,
(+,-), is at least 13 standard deviations away from the best fit
of the incorrect sign combinations. The fits for the different sign
combinations are shown in figure \ref{invMassDist}. 

The log likelihood fit for the mass differences yields 
$M_{\nino{2}}-M_{\nino{1}}=58.8GeV \pm 0.3$ and
$M_{\nino{3}}-M_{\nino{1}}=82.25GeV \pm 0.2$. The generator
level values for these masses are $58.6$ GeV, and $82.3$ GeV.

 It should be noted that theoretically we could do the same study with the 
invariant mass of the hadronic $Z$ decay. However, reconstruction of 
di-lepton mass distribution will have much less uncertainty than 
reconstruction of the di-jet mass distribution. Also the efficiency and 
sculpting of the di-jet distribution is much more complicated than for
the di-lepton due to the jet algorithms.

\subsection{LSP Mass From Energy Distribution}

The invariant mass distribution of the two leptons had very weak
dependence on the LSP mass. However, the distribution of the energy of the
two leptons
will depend on the LSP mass. It will also depend on the $\nino{2}$ and 
$\nino{3}$ masses, the theoretical invariant mass distribution of
the two leptons, the beam energy, the spectrum of initial state
radiation, and the spin correlations in the neutralino decays. 

The energy distribution is numerically calculated by integrating over all
of the above mentioned distributions. 
It should be noted that the version of Isajet
used in this analysis does not calculate the spin correlations of
the neutralino decays
\cite{hbearPCom}. The spin correlations of the neutralino decays
 affect the angular distribution of
the virtual $Z$ in the rest frame of the parent neutralino along the boost
direction. This angular
distribution is flat in the Monte Carlo because of the lack of neutralino
spin 
correlations. This translates to a flat energy distribution in the lab frame 
for a given value of di-lepton invaraint mass and beam energy. 

\begin{figure*}[t]
\centering
\includegraphics[width=135mm]{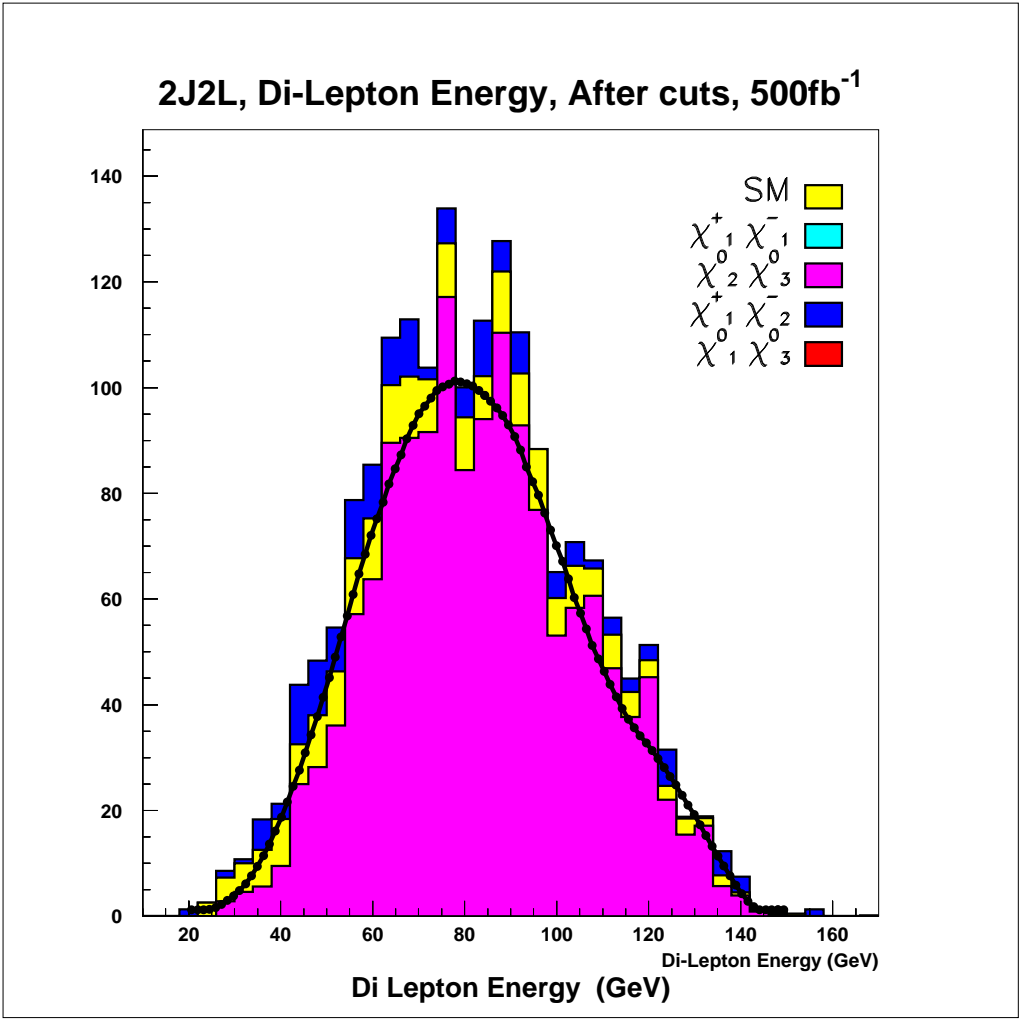}
\caption{This shows the energy distribution of the two leptons 
in the 2 lepton 2 or 3 Jet and missing energy decay signature. The magenta
color is the signal, the yellow is the contribution from standard
model backgrounds and the blue is background from chargino production.
The curve on the plot is the calculated energy distribution for the signal. 
} \label{edistplot}
\end{figure*}

The data and the calculated energy distribution are shown in figure
\ref{edistplot}. The standard model background distribution is fit 
to an order 3 polynomial and is combined with the calculated signal 
distribution for a log likelihood fit to the data. In the fit the $\nino{1}$
(LSP) mass is allowed to vary, and the mass differences of the $\nino{2}$
and the $\nino{3}$ with the LSP mass are allowed to vary within 3 standard
deviations of the measured mass differences obtained from the invariant
mass distribution fit. The fit finds an LSP mass of 
$108.3 \textrm{ GeV} \pm 1.0$ GeV. The generator level value of the LSP mass
is $107.7$ GeV.

\section{Conclusions and Future Directions}

We have presented preliminary results for the expected uncertainties that a
500 GeV linear collider would have in measuring mass and cross sections 
from $\nino{2}\nino{3}$ production in a focus point region using the 
2 lepton plus 2 or 3 jet plus missing energy detector signature. We have found
that for 250 $\textrm{fb}^{-1}$ of left polarized data and 
250 $\textrm{fb}^{-1}$ of right 
polarized data we could determine the cross sections to within a 
uncertainty of 4\% and 5\% respectively. By using the invariant mass
distribution of the two leptons 
of the combined 500 $\textrm{fb}^{-1}$ data set we could measure the 
mass differences $M_{\nino{2}}-M_{\nino{1}}$ and $M_{\nino{3}}-M_{\nino{1}}$
with $0.3$ GeV and $0.2$ GeV uncertainty respectively. By using the energy
distribution of the two leptons we can measure the $\nino{1}$ (LSP) mass
with an uncertainty of $1.0$ GeV. 

Other work currently being done by the Cornell/Florida LC Cosmology Group 
is the study of $\nino{1}\nino{2}$ production, $\cinop{1}\cinom{1}$ production,
and $\cinop{1}\cinom{2}$. There is also work being done to convert the
uncertainties in mass and cross section measurements into uncertainties 
in the relic density. If the uncertainty in the relic density of
the LSP is small enough, we can compare our prediction to WMAP constraints
and determine whether or not the LSP actually is the Dark Matter.

\begin{acknowledgments}
The authors wish to thank M. Peskin for his many valuable insights and support.

We would also like to thank T. Barklow for providing our Standard Model 
backgrounds. 
\end{acknowledgments}

\end{document}